%% file: 0.0-main.tex
\documentclass[sigconf]{acmart}
\usepackage[utf8]{inputenc}
\AtBeginDocument{%
  \providecommand\BibTeX{{%
    \normalfont B\kern-0.5em{\scshape i\kern-0.25em b}\kern-0.8em\TeX}}}

\setcopyright{none}
\copyrightyear{2021}
\acmYear{2021}
\acmDOI{}
\acmConference[Pending Publication]{Pending Publication}{April 12}{2021}
\acmPrice{}
\acmISBN{}
\thanks{LA-UR-21-23448}
\thanks{The U.S. Government retains for itself, and others acting on its behalf, a paid-up nonexclusive, irrevocable worldwide license in said article to reproduce, prepare derivative works, distribute copies to the public, and perform publicly and display publicly, by or on behalf of the Government. The Department of Energy will provide public access to these results of federally sponsored research in accordance with the DOE Public Access Plan. http://energy.gov/downloads/doe-public-access-plan}
\title{Designing a scalable framework for declarative automation on distributed systems} 

\author{J.~Lowell Wofford}
\affiliation{%
   \institution{Los Alamos National Laboratory}
   \city{Los Alamos}
   \state{New Mexico}
}

\date{\today}

\newcounter{reqs}
\refstepcounter{reqs}
\newcommand{\req}[3]{\item [(R\thereqs) #2] #3\label{req:#1}\refstepcounter{reqs} }
\usepackage{listings}
\newcommand{\prol}[1]{\lstinline[language=Prolog,basicstyle=\ttfamily,keepspaces=true]!#1!}
\begin{document}

\begin{abstract}
\input{0.1-abstract.tex}
\end{abstract}

\keywords{high-performance computing, distributed computing, operating systems}

\maketitle

\ccsdesc[500]{Information systems~Computing platforms}
\ccsdesc[500]{Computer systems organization~Distributed architectures}

\section{Introduction}
\label{section:introduction}
\input{1.0-introduction.tex}

\section{Theory and requirements}
\label{section:anatomy}
\input{2.0-anatomy.tex}

\section{The Kraken framework}
\label{section:framework}
\input{3.0-framework.tex}

\section{Problem constraints and use caes}
\label{section:problem-constraints}
\input{5.0-problem-constraints.tex}

\section{Conclusion}
\label{section:conclusion}
\input{6.0-conclusion.tex}

\appendix
\input{reqs.tex}
\section*{Acknowledgements}
This work was supported by the U.S. Department of Energy through the Los Alamos National Laboratory. Los Alamos National Laboratory is operated by Triad National Security, LLC, for the National Nuclear Security Administration of U.S. Department of Energy (Contract No. 89233218CNA000001).

\bibliographystyle{ACM-Reference-Format}
\bibliography{bibliography}

\end{document}

%% file: 0.1-abstract.tex

As distributed systems grow in scale and complexity, the need for flexible automation of systems management functions also grows.  We outline a framework for building tools that provide distributed, scalable, declarative, modular, and continuous automation for distributed systems.  We focus on four points of design: 
1) a state-management approach that prescribes source-of-truth for configured and discovered system states; 
2) a technique to solve the declarative unification problem for a class of automation problems, providing state convergence and modularity; 
3) an eventual-consistency approach to state synchronization which provides automation at scale; 
4) an event-driven architecture that provides always-on state enforcement.

We describe the methodology, software architecture for the framework, and constraints for these techniques to apply to an automation problem.  We overview a reference application built on this framework that provides state-aware system provisioning and node lifecycle management, highlighting key advantages.  We conclude with a discussion of current and future applications.

%% file: 1.0-introduction.tex

The ongoing growth in system scale and complexity of clustered and distributed systems has lead to complexity in administration and management of large systems.  There have been many approaches to dealing with system management complexity, ranging from ``hire more systems administrators'' to ``automate all the things.''  Unfortunately, both of these approaches often hit walls.  Already, leadership-class HPC systems have scaled to tens of thousands of independent compute nodes, and it is expected that these numbers will continue to grow as we reach the exascale era.  Even still, these numbers are eclipsed by those in the web-scale and cloud space.  We have already reached scales where ``hire more systems administrators'' is untenable.

While adding administrators hits a wall at scale, automating all the things tends to hit a wall at complexity.  There are many existing efforts to automate large-scale systems, ranging from configuration management and orchestration solutions like Ansible\cite{ansible} and Puppet\cite{puppet}, to container orchestration solutions like Kubernetes\cite{kubernetes} and Docker Swarm\cite{docker}.  In most of these cases, sufficiently complex environments make automation difficult.  There are many pitfalls to automating at scale.  For instance, it is easy for large scale systems with many administrators to eventually have conflicting automation programs: more than one workflow trying to automate the same component in the same way.  While configuration management driven systems often can control many aspects of a system, they do not perform ``live'' automation; they simply enforce state (up to the level of their specification) when they are told to run.

In this paper, we propose a theory and framework of automation specifically aimed at the management of distributed systems.  We examine some of the pitfalls that existing automation tools often hit, and propose a methodology for a novel approach to system automation.  This methodology was developed hand-in-hand with a software implementation, Kraken\cite{kraken}, which implements most of the features we will outline in this paper; however, we believe that the methodology supersedes the particular implementation. 

\subsection{A case for declarative automation at scale}
\label{subsection:a-case-for-declarative-automation-at-scale}

When discussing system automation, one can draw a distinction in approaches between declarative automation and imperative automation.  In theory, any task that can be done with one of these can be done with the other, and the choice between them, as a quick internet search will reveal, is largely a matter of preference.  The declarative vs.~imperative debate has become highly contended.  We do not intend to settle that debate in this paper.  Rather, argue that declarative automation provides some strong benefits to combat some of the common challenges that occur in very large and very complex systems.  

The distinction between declarative and imperative automation hinges on how the administrator of the automation workflow interacts with the automation system.  In a declarative system, the administrator's task is to ``declare'' what the state of the system should be; the automation systems task is to translate this desired state into a set of actionable tasks and perform them, ultimately ``converging'' to the point where the system is in the desired state.  The theory of declarative systems dates back to the early days of computer science, and elements of declarative systems can be found in technologies ranging from the programming language Prolog to the Puppet configuration management system.

In an imperative system, the administrator's task is to provide a sequence of actions that should be applied to systems meeting a set of criteria.  The automation system's job is to make sure that these actions are applied when the criteria are met.  The imperative approach can be said to apply to anything from a basic shell script to a complex Ansible playbook.

Both of these approaches have advantages.  There are several reasons one might choose the imperative appraoch.  In many ways, it is the straight-forward and ``usual'' way to automate systems with shell scripts; this is a skill most systems administrators already possess.  Imperative automation flows can also be highly scalable.  In many cases, imperative flows can be applied in parallel to vast numbers of systems at once with little effort.  

Declarative systems can be difficult to scale.  This is largely owing to their state-driven design which will be discussed at length in this paper.  However, declarative systems bring strong advantages when dealing with complexity.  We call attention to four strong advantages: 
\begin{enumerate}
    \item Abstracting implentation details from adminstrative tasks allows administrators to focus on \emph{what} needs to be done rather than \emph{how} to do it.
    \item An event-driven declarative system can efficiently provide ``continuous'' automation.
    \item A sufficiently defined event-driven declarative system can be self-healing.
    \item Declarativesystems are naturally state-aware, allowing them to make context-based decisions, like when it is appropriate to perform a desctructive action.
\end{enumerate}
These advantages will be demonstrated throught this paper. There are many other points that could be made in favor of this approach, but we believe these four features, when demonstrated, are enough to motivate this approach for large-scale and complex systems. 

\subsection{Design objectives}
In order to achieve these advantages, we distill five guiding design objectives:
\begin{description}
    \item[Distributed] We need to be able to auotmate systems with many components.
    \item[Scalable] We need to be able to automate system functions at scales up to and beyond 10's of thousands of nodes.
    \item[Declarative] Automation actions should be based on declarations of desired system states rather than procedures for reaching system states.
    \item[Modular] The tooling to reach system states should be modular; it should be possible to reach desired system states in different ways for different systems or components.  Additionally, the states tracked and managed should be extensible.
    \item[Continuous] Automation processes should continue to operate through the lifecycle of the system.  If a component deviates from the desired state, it should be, within reason, corrected without intervention.
\end{description}

\subsection{Overview and scope}

The remainder of this paper will discuss an approach to scalable, declarative, modular, and continuous automation.  We will begin with a high-level discussion of these design objects and what they require programmatically.  We will do so initially through an analogy to declarative programming.  This will allow us to identify a set of core requirements that we must meet.  We will then be in a position to describe the Kraken framework, its approach to resolving these challenges, and the overall architecture of a Kraken-based application.  Finally, we will discuss the limitations and uses of the Kraken framework.

This paper is not intended to be an exhaustive survey of the broad fields of automation, declarative logic or systems management.  Rather, the focus is on a specific design pattern that brings strong advantages to large-scale and complex systems management.  Further, we will use many examples throughout that are focused on the management of node lifecycle within distributed systems, but we believe this same approach can be applied to many other applications.  

%% file: 2.0-anatomy.tex

We have established a set of design requirements for the kind of automation we would like to perform. Namely, we want a way to achieve scalable, declarative, modular, and continuous automation.  We will now examine some of the problems that arise from these requirements.  We will use an analogy to declarative programming to discuss some methods that can solve these problems and establish a set of core features that our framework must have to meet our design requirements.  

\subsection{An analogy to declarative programming}
\label{section:anatomy-of-declarative-automation}
We begin by examining in more detail what we mean by ``declarative automation'' and what the anatomy of such a framework might look like by learning from known methods of declarative programming.  Later, we will discuss extending this automation to be distributed, modular, scalable, and continuous.

\subsubsection{A very rapid survey of declarative programming}
\label{subsubsection:declarative-programming}

We can derive many of the concepts of declarative automation from the well-studied field of declarative programming.  Declarative programming is often discussed in terms of Prolog\cite{warren1977prolog,lloyd1994practical}, being the paradigmatic example of a declarative programming language.  We will adopt some of that language.  Borrowing some high-level concepts and terminology from declarative programming will help derive language, methods, and requirements for declarative automation.

\sloppy We begin with the Kowalski equation\cite{Kowalski_1979}:  \prol{Algorithm = logic + control}. Kawalski asserts that a program consists of logic statements combined with control statements.  In this context, logic refers to expressions with logical operators, such as equivalence (``\textbf{if} $X$ is a sea monster...''), while control relates to execution control statements, like ``jump to line 6.'' The Kawalski equation provides a generically accurate prescription for a program to run on a Turing machine.

The declarative programming model eliminates the control portion of the Kowalski equation. A declarative program is constituted strictly of logic, which is why it is sometimes referred to as \emph{logic} programming.  In the language of Prolog, a program consists of \emph{facts}, \emph{rules}, \emph{goals}, and \emph{queries}.  We'll illustrate these concepts through an example program that alludes to power control through IPMI.
\begin{description}
	\sloppy\item [Facts] are the particular states of the program, e.g.~\prol{ipmi(system1).} (``system1 is ipmi'')
	\sloppy\item [Rules] establish a set of logic conditions for the program, e.g.~\prol{ipmi_poweron(X) := ipmi(X).} (``ipmi => ipmi\_poweron'')
	\item [Goals] establish the truth of a predicate and a set of facts, e.g.~\prol{ipmi_poweron(system1).} (evaluates to true)
	\item [Queries] queries can be run against a set of facts and rules, e.g.~\prol{ipmi(X).} (``X where X is ipmi'').  Results in \prol{system1.}
\end{description}
Putting these together, this makes a runable program:
\begin{lstlisting}[basicstyle=\ttfamily,columns=fullflexible,keepspaces=true,escapechar=ß,frame=single,label={lst:power},caption=Representative Prolog for abstracted system power]
ipmi(system1).
redfish(system2).
ipmi_poweron(X) :- ipmi(X).
redfish_poweron(X) :- redfish(X).
poweron(X) :- ipmi_poweron(X) ; redfish_poweron(X).ß\footnote{In Prolog, $;$ is the \emph{or} operator.}ß
/* goal */
:- poweron(system2) /* true. */
\end{lstlisting}
The entire program consists of predicate logic.  There are no \emph{control} statements that specify procedural actions, like ``iterate through this list and test equivalence,'' when evaluating a query or goal.

Prolog may appear different than the explanation of declarative automation given above.  We described a declarative automation process as one in which we describe a final state, and a sequence of actions converges the actual state to it. Prolog, on the other hand, evaluates goals and queries on existing facts and propositions.  The two are not equivalent, but one can establish a useful analogy between them.  We can use this as a rubric to map Prolog terms onto the declarative automation problem:
\begin{enumerate}
	\item The \emph{facts} in our problem are the \emph{states} of the system.  E.g.~the power state of a system, whether an application is running, the available memory, etc.
	\item The \emph{rules} become idempotent \emph{actions} on a system.  E.g.~power on, run an application.  We call these combinations of rules and actions \emph{state mutations} since they change the state of a system to meet a goal rather than only evaluating a system's state.
	\item \emph{Goals} map to \emph{desired states} of the system, E.g., the system is powered on, or an application is running. 
\end{enumerate}
An automation program consists of a set of system states, a set of mutations, and a goal statement. The goal evaluates to true if a sequence of mutations, i.e.~a \emph{mutation chain}, can be derived that make the goal true.

The process of establishing this sequence of mutations shares a logical relationship with the rules.  
Consider that Listing \ref{lst:power} would evaluate to \prol{true.} because there is a logical deduction through \prol{redfish_poweron(system2)}.  If we associate an \emph{action} to \prol{redfish_poweron(X)} to turn on the system's power (idempotently), this will achieve powering on \prol{system2}, and it will do so without requiring knowledge of \emph{how} to control \prol{system2}'s power.

Through this analogy, we can describe automation processes that meet some of the objectives we prescribed above.  We intend to provide a specification for our system's state and have that state achieved without giving a specific procedured.  We can, therefore, discuss some of the challenges of declarative automation with the well-studied tools and language of declarative programming.  

We mentioned that declarative programming omitted control statements while also asserting that control statements are required to run on a Turing machine.  Prolog programs run on Turing machines and do not specify control statements; therefore, Prolog control statements must be implicit.  The control statements in Prolog come from the algorithms for logical deduction within Prolog.  The utility in our analogy to deductive programming is in our ability to learn from the deductive process for queries and goals and what this process implies for specifying and using rules (mutations).

Without going into the implementations of these mechanisms yet, we can outline three requirements we must meet to achieve declarative automation:
\begin{description}
    \req{state-mgmt}{State management}{We must possess a representation of the current state of the system along with the tooling to track, evaluate and report on these states.}
    \req{state-spec}{State specifications}{We must possess a mechanism for defining and asserting goals, or desired states,  for the system.}
    \req{state-conv}{State convergence}{We must possess a mechanism to find and actuate a sequence of state mutations (a \emph{mutation chain}) that will achieve the desired state.}
\end{description}
These three pieces constitute the core functionality that an automation engine must possess to be declarative.  We will explore each of these in turn.

\subsubsection{State management: event-driven facts and continuous automation}
\label{subsubsection:state-management}

System state management is at the heart of our approach to system automation.  The system state provides the facts for our declarative automation.  In an idealized declarative program, the facts are static entities.  As long as this remains true, a declarative program is side-effect-free.  Nothing should modify the program's facts outside of the program.  Being natively side-effect-free has advantages for the automation workflow.  Importantly, side-effect-free execution allows for efficient distribution of execution.  If all of the state information of an automation flow remains within the local scope, we can parallelize the automation without state synchronization overhead.

In any practical system the facts of the system are event-driven.  For instance, the power state of a system changes when someone presses the power button. This changed power state is an event outside of the program's scope.  We refer to these event-driven facts as \emph{discoverable state}, as they represent the current state of the systems discovered through inspection.  In a clustered system, a node might fail and the discoverable state of that node changes.  These unexpected events constitute side-effects in the execution of automation programs, and we must handle them carefully.  

The nature of the declarative system allows us to handle discoverable facts robustly.  When a state changes unexpectedly, the entire automation flow can be interrupted and re-evaluated with the updated discoverable state.  If the state change necessitates a different mutation chain, that chain will begin to execute.  If it does not, it will naturally resume its progress.   This approach allows the system to quickly adjust its behavior to match changing states due to outside, asynchronous events while only having a minimal impact on concurrency by keeping executing workflow state local.  

\begin{figure}[htb]
	\begin{center}
	\includegraphics[width=\columnwidth,angle=0]{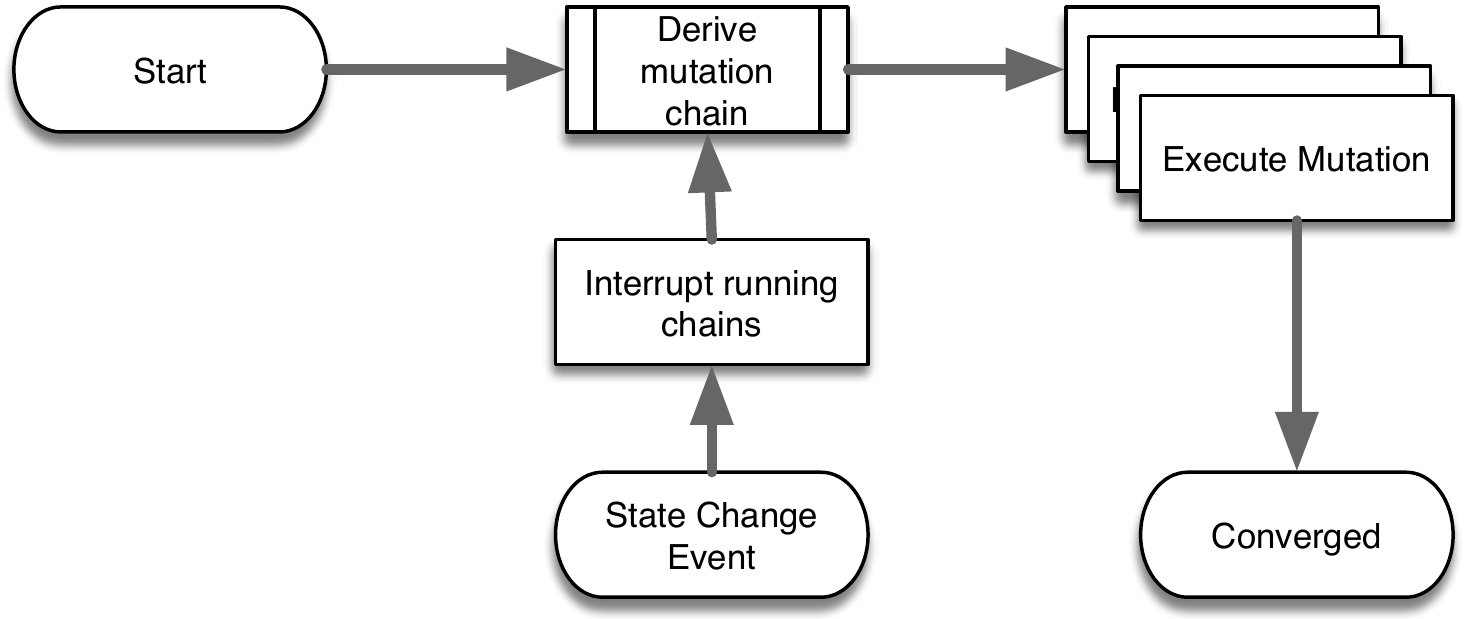}
	\caption{Event driven design for continuous automation.}
	\label{fig:event-driven}
	\end{center}
\end{figure}

The preceding also suggests how to make automation continuous.  By adopting an overall event-driven design and keeping the automation program persistent, the same process that handles events while executing a mutation chain can also manage events after completing the mutation chain.  These events can be processed in the same way as described above, potentially starting new mutation chains as needed.  For instance, if an automation program requests that a process run and completes that request, but later the process exits, it would generate an event that would re-start the mutation chain for starting that processd.  This approach leads to continuous automation and--within confines of what mutations the program is capable of--self-healing capabilities.  Additionally, changes in state specification (goals) can trigger the same re-evaluation process, providing the ability to alter running systems.

Based on this discussion, we can add some more requirements for the framework:
\begin{description}
	\req{discoverable}{Discoverable}{States which have a source of truth outside of the program must have a means of inspecting and updating the local representation of the states.}
	\req{event-driven}{Event-driven}{To achieve continuous operation and reliable execution, changes in discoverable state or state goals should be handled as asynchronous events.}
\end{description}

\subsubsection{State specifications: how to specify queries}
\label{subsubsection:state-specifications}

The discoverable state is only half of the state information we require.  We also need to know the state we wish to be in, that is, the system's state goals.  This is a kind of configuration, so we refer to it as \emph{configuration state}.  Generally, this information will be provided directly (via an API) or indirectly (via an intermediate program) by the automation program user.  As we noted in \ref{subsubsection:state-management}, we do not assume that this specification remains static, and we leverage event-driven design to update the system on a configuration state change dynamically.

We add the following requirement to the list:
\begin{description}
	\req{cfg-states}{Configuration states}{We must provide a mechanism for users or programs to set and update configuration states. Note that this is related to but distinct from (R\ref{req:state-spec}).}
\end{description}

\subsubsection{State convergence: Unification \& mutation chains}
\label{subsubsection:state-convergence}
\begin{lstlisting}[float=*,language=Prolog,basicstyle=\ttfamily,columns=fullflexible,keepspaces=true,escapechar=ß,frame=single,label={lst:power2},breaklines=true,postbreak=\mbox{\textcolor{red}{$\hookrightarrow$}\space},caption=Prolog power control on two platforms]
/* Facts */
power(off, system1).
platform(redfish, system1).
power(on, system2).
platform(ipmi, system2).
redfish_power(on, Node) :- platform(redfish, Node), power(off, Node). 			/* Rule 1 */
redfish_power(off, Node) :- platform(redfish, Node), power(on, Node). 			/* Rule 2 */
ipmi_power(on, Node) :- platform(ipmi, Node), power(off, Node).		  			/* Rule 3 */
ipmi_power(off, Node) :- platform(ipmi, Node), power(on, Node).		  			/* Rule 4 */
power(Status, Node) :- redfish_power(Status, Node) ; ipmi_power(Status, Node).	/* Rule 5 */
\end{lstlisting}
In Section \ref{subsubsection:declarative-programming} we introduced the concept of state convergence (R\ref{req:state-conv}) and the declarative deduction process.  As we noted, the control logic for a procedural program comes from the deduction algorithm.  The same is true in our implementation of declarative automation.  Specifically, state convergence relies on our ability to compare a discoverable state and a configuration state specification and deduce a mutation chain that will converge the two to equivalence.

The primary tool in procedural programming for deducing a goal or query from a set of facts and rules is \emph{unification}.  Unification binds a value(s) in the query or goal to create a new rule.  It will continue to do so recursively until it arrives at a true or false fact.  Consider the evaluation of the goal \prol{poweron(system2)} from Listing \ref{lst:power}.  The result of \prol{true.} is achieved through a sequence of unifications like this:
\begin{enumerate}
	\item Bind \prol{system2} to \prol{X} to derive the unification rule \prol{ipmi_poweron(system2) ; redfish_poweron(system2).}
	\item Bind \prol{ipmi_poweron(system2)} to \prol{ipmi{system2}}.
	\item Notice that \prol{ipmi(system2)} is not a fact, so this evaluates to \prol{false.}
	\item Backtrack to the branch in \prol(poweron(X)), evaluate \prol{redfish_poweron(system2)}
	\item Bind \prol{redfish_poweron(system2)} to \prol{redfish(system2)}.
	\item Notice \prol{redfish(system2)} is a fact, therfore \prol{true.}
	\item This registers the fact \prol{poweron(system2)}, and evaluates \prol{true.}
\end{enumerate}
This program is a simplistic example with only one search depth layer, but we see that the deduction process is a depth-first search backward, where we accumulate rules and facts through unification.  

Now let's see how we can apply the same deduction concept to the automation case.   We will use a slightly more sophisticated program (See Listing \ref{lst:power2}). This example starts looking more like a realistic way we might think of states of nodes.  We have two kinds of state, \prol{power}, and \prol{platform}.  Each system has a value for each of these kinds of states.   

Recall that, for our automation, rules have a special role.  They get bound with actions to make mutations, so we are especially interested in which rules get used.   Let's consider an example goal evaluation.  
If we ``power on'' \prol{system1} with \prol{power(on, system1)}, it will use rules in this sequence: \prol{Rule 5 -> Rule 1} and evaluate \prol{true.}.  The other available path through \prol{Rule 3} would evaluate \prol{false.}, but we never even have to evaluate that path.  

There are three observations we can make from this example.  First of all, we used two rules in the first goal evaluation.  Rule 1, when bound to an action, would presumably turn on the node power using the Redfish mechanism.  Rule 5 acts as a logical abstraction between the two platforms.  If we were to procedurally power on this system, the order of operations is reversed from this deductive order; that is, we need to proceed from the order of inference, not the order of deduction.  

Second, there is a difference here in how rules work and how mutations work, and this is a point at which the Prolog analogy breaks down.  We successfully evaluate all three goals, but none of these goals alter the facts; they evaluate true because a path exists.  When we evaluate power-on in goal 1, it doesn't change the facts such that \prol{power(off, system1)} is \prol{false}.  This is distinctly different than the case of automation.  In automation, the mutation associated with \prol{power(on,system1)} would alter the discoverable state of \prol{system1} on success, and subsequent mutations would depend on that discoverable state being reached.  Implicitly, a specification of which states a mutation is expected to alter is a fundamental attribute of a mutation.

Finally, the structure of mutation chains depends on a set of criteria assigned to rules.  If we evaluate \prol{power(off, system2)} instead we arrive at a path through Rule 4 precisely because that rule has the correct set of \emph{requirements} associated with it, namely that \prol{power == on} and \prol{platform == ipmi}.  This suggests that a mutation should include a specification for conditions under which it is applicable.  

Combining these inferences, we see that a mutation consists of a least the following components: 1) A specification of what state values the mutation is capable of altering, e.g.~\prol{power = off} to \prol{power = on}; 2)  A specification of requirements that must be met for the mutation to act, e.g.~\prol{platform} must be \prol{redfish}; 3) An association with a procedural action to take, e.g.~turn on the power.

A mutation chain is then a sequence of actions that will alter the state of a system in steps to converge the discovered and configured state.  The mutations in the chain in conjunction with their associated actions provide a procedural recipe--the control statements--of the automation program.

\sloppy We can define a version of the unification operation for mutations that will allow for an analogous procedure for the deduction of these mutation chains from a set of mutations. Whereas unification is the binding of a rule with a fact to arrive at new rules and facts, the unification of mutations is the binding of a set of states with a mutation to arrive at a new set of states.  For instance, the unification of the \prol{redfish_power(on, X)} mutation with the set of states \prol{power(off, system1), platform(redfish, system1)} is the new set of states \prol{power(on, system1), platform(redfish,system1)}.

Finally, we can prescribe an abstract approach to finding mutation chains.
\begin{enumerate}
	\item Starting with the desired state
	\item For each mutation, unify to create a new state
	\begin{enumerate}
		\item If the new state is equivalent to the discoverable state, the inverse of this path is the mutation chain.
		\item If the new state has already been seen, stop the recursion
		\item Recurse to 2
	\end{enumerate}
\end{enumerate}
This searches the space of available states for a path that leads from the desired state to the current state.  This path's inverse is the mutation chain.  Note that depending on constraints applied and the structure of mutations, this may not be deterministic or terminating; this is considered left to the implementation.

We can introduce an additional requirement:
\begin{description}
	\req{mutations}{Mutation management}{We must provide a mechanism for finding and executing chains of state mutations.  Note that this is related to but distinct from (R\ref{req:state-conv})}.
\end{description}

\subsection{Distributing state}
\label{subsection:distributing-state}

The requirements as we have described them so far will create an automation engine that operates only locally.  We will need to move beyond the declarative programming analogy to examine this objective.  One of the stated design objectives, and indeed one of this project's core objectives, is to build automation tools that can work across large distributed systems.  We assume that we have an extensive collection of systems with disparate resources that can communicate over a network.  We further assume that each of these systems' desired states can be described with the same sets of state variables.  We wish to distribute the automation capabilities we have described across this set of systems.

In the context of an HPC system, we may have many systems. Each system needs to run some higher-level function like compute-enabled system images or I/O-routing capabilities.  We can describe all of the nodes' desired functions with a common set of state variables, e.g.~which images and applications should be loaded.  Once the state is appropriately distributed, the automation program can run locally to ensure that the local desired states are converged.  In this sense, the automation is distributed, i.e.~automation programs run in a distributed way on localized state. 

Before we can distribute automation, we must know how to parallelize automation.  Given the event-driven design of (R\ref{req:event-driven}), we already operate in a mode where a local execution context can treat the state as having local scope until execution is interrupted due to a raised event.  As was briefly discussed in Section \ref{subsubsection:state-management}, different processes can autonomously execute until events are raised.  The event-driven model then provides for parallel execution of automation.  

Given this prescription for parallel execution, distributing execution is relegated to synchronizing state and providing a local execution environment.  There is no one answer for how this is to be done.  The methods used to synchronize state and the consistency guarantees that synchronization provides will impact what kinds of automation are allowable with the particular implementation.  We will leave the requirements more generic for now:
\begin{description}
	\req{sync}{State synchronization}{We must provide a mechanism to synchronize state across system nodes.}
	\req{local-execution}{Local execution}{We must provide a local automation execution environment on nodes.}
\end{description}

\subsection{Achieving modularity}

We have yet to address the objective of modularity.  However, we have laid requirements out in such a way that modularity is natural.  The automation framework we have described so far makes no assumptions about what kinds of actions mutations take, leaving the types of automation actions to be performed generic.  We also noted in Section \ref{subsubsection:state-convergence} that the determination of mutation chains is dependent on which mutations are present.  Finally, in the same section, we noted that mutations consist of at least: 1) which states they modify; 2) requirements to execute; 3) and a linked action.  

If we allow modules to provide the mutations, the automation actions we perform, including the chains we perform them in, will depend on the set of modules present.  We hinted at this capability with Listing \ref{lst:power2}.  We can envision different modules providing the mutations for \prol{IPMI} power control and the mutations for \prol{Redfish} power control.

Modules must perform at least one other function.  Recalling that discoverable states are discovered through inspection, if we are to make the framework modular, modules must also provide this inspection process.  For instance, modules would discover power states in the example above.

Modules may take on other roles, such as providing microservices or other utilities, but we can assign two additional requirements:
\begin{description}
	\req{mod-mut}{Module mutations}{To achieve modularity, state mutations should be provided by modules.}
	\req{mod-disc}{Module discoveries}{To achieve modularity, discoverable states should be assigned by modules.}
\end{description}

%% file: 3.0-framework.tex

\subsection{Architecture overview}
\label{subsectinon:architecture-overview}

Up until this point, we have examined in a generic context how to create an automation engine that is distributed, scalable, declarative, modular, and continuous.  We broke out a set of requirements, terminology, and methods.  We will now turn to describe the Kraken framework.  The Kraken project provides a programming framework to build tools that meet the requirements of the previous section and the overall design objectives.  We want to emphasize from the beginning that, as a framework, Kraken itself performs no automation.  As we will see, Kraken, with the addition of \emph{modules} and \emph{extensions} builds tools that perform specific automation tasks.  For this reason, we make a clear distinction between Kraken and Kraken-based tools.  In this section, we will examine the architecture of this framework.

Unlike the last section's generic automation discussion, when designing Kraken, we made certain assumptions and optimizations that keep Kraken tools scalable and robust.  These optimizations can also limit what can be implemented in Kraken, and we will discuss those limitations more in a later section (See Section \ref{section:problem-constraints}).  We will be careful to highlight throughout where and why we made these choices.

\subsubsection{A top-level view}

Kraken is written in Go\cite{golang}, and while it is expected that Kraken-based tools are also written in Go, it is possible for Kraken tools to interact fluidly with programs written in other languages through the use of open and portable APIs.   Kraken relies heavily on Protobuf\cite{protobuf} for data serialization and gRPC\cite{grpc} for internal communication.  Protobuf and gRPC allow us to make the Kraken APIs language-portable and efficient.  In addition, Kraken includes core modules (see discussion on modules) that expose an OpenAPI 2.0 (Swagger) Restful API as well as WebSocket support for event streaming for any Kraken tool that wishes to use them.

Kraken uses a tree-like method for state distribution\footnote{Kraken tools can provide automation for single-host instances too, but we will assume throughout that we are running in a distributed way.}.  Generally, users will interact on the parent level to control automation actions on the child level.  Each node--parents and children--must be running a copy of the Kraken tool.  Parent Kraken tools can also perform actions on children, even before they are running local instances.  For this reason, Kraken mutations also must define the context (\emph{child} vs. \emph{self}) that they are intended to be run in.  The Layercake example (See Section \ref{section:problem-constraints}) uses this functionality to provision children from bare-metal before controlling them.

A Kraken-based tool usually consists of at least the following three components:
\begin{description}
	\item [Modules] A set of modules that define the automation flows for the tool.  These modules provide mutations (detailed in section \ref{subsection:mutations}), routines to discover state values, and potentially arbitrary other functions, like providing system services.  Modules are linked to the tool at compile time.
	\item [Extensions] Most tools will need to extend the base set of state variables that are tracked. Extensions allow the state schema to be extended to include new values.
	\item [Entry point] A program entry point must be included that takes the necessary steps to initialize the tool and start the core Kraken components.
\end{description}
Each of these components can be partially generated by an included tool called \prol{kraken-builder} when combined with YAML-based description files.

Any tool that is built from these components generates a single executable.  The executable must be run on any node participating in automation.  The executable contains all of the pieces necessary to implement any automation flow that is defined through the mutations provided by the included modules.  
The process for designing a Kraken-based tool can be outlined as:
\begin{enumerate}
	\item Decide which components the tool should control, e.g.~running container images on nodes.
	\item Enumerate the actions that can be taken on these components, e.g.~attach image, run a container, stop the container.  These actions, combined with handlers for performing them and a set of requirements for running them (e.g.~node must be in a ready state), define the mutations.
	\item Decide which bits of state information are required to perform actions, e.g.~name of the container, location of the container, desired state of the container.  Build an extension that defines these state variables.
	\item Now, build module(s) that implement the mutations and discover state values (and possibly other things).
	\item Create an entry point application that links these modules and extensions.
\end{enumerate}
At this point, the application can be built and run.  However, generally, the tool will not perform any function until a configuration state is provided.  State injection can be done at any time through the Restful API, the ModuleAPI (gRPC), or by providing a state definition file on runtime.  State injection is usually performed through a parent node.

\subsubsection{The five engines of Kraken}
Kraken implements its core functionality through five ``Engines'' that run and actively handle internal events.  These engines run concurrently as Go-routines, and handle tasks through event handlers.  Each engine performs functions that address different pieces of the requirements.  

\paragraph{EventDispatchEngine}  The EventDispatchEngine (EDE) handles event management within Kraken.  Events, which take the form of Protobuf defined objects, can be triggered for many different reasons.  For instance, when the StateDifferenceEngine detects that a state has changed, a \prol{STATE_CHANGE} event is emitted.  When the StateMutatonEngine would like to request a mutation, a \prol{STATE_MUTATION} event is emitted.  These events flow to the EDE through Go channels, where they are processed and dispatched to other components.  Any component within Kraken, including modules (via gRPC), can subscribe to events by type and by filter parameters.  Any component can also emit events into the EDE to be handled by other components.

The EDE is the simplest of the Kraken engines, but it is critical to understanding how data flows with Kraken. Events trigger almost every action that the Kraken framework performs.   The EDE is also the core component to meeting the (R\ref{req:event-driven}) (Event Driven) requirement.  

\paragraph{StateDifferenceEngine} The StateDifferenceEngine (SDE) is the primary engine involved with internal state management (R1).  States in Kraken are stored in Protobuf objects.  The base state object is a \prol{Node}, which contains only a few state variables used internally within Kraken.  The \prol{Node} objects are extended through \emph{extensions}.  A state store consists of a collection of \prol{Node} objects, which are keyed by a unique UUID.  The SDE maintains two copies of every node object, representing the discoverable state and the configuration state.  Generally, the contents of a \prol{Node} object are thought to pertain to a physical node.

The SDE performs three primary functions:
\begin{enumerate}
	\item The SDE provides the (in-memory) storage for the Kraken state store. 
	\item The SDE provides a simple query language for accessing and updating state information.  It also can update discoverable information by listening for \prol{DISCOVERY} events.
	\item The SDE tracks changes to state and emits \prol{STATE_CHANGE} events when state changes occur.
\end{enumerate}

\begin{figure}[htb]
	\begin{center}
	\includegraphics[width=\columnwidth,angle=0]{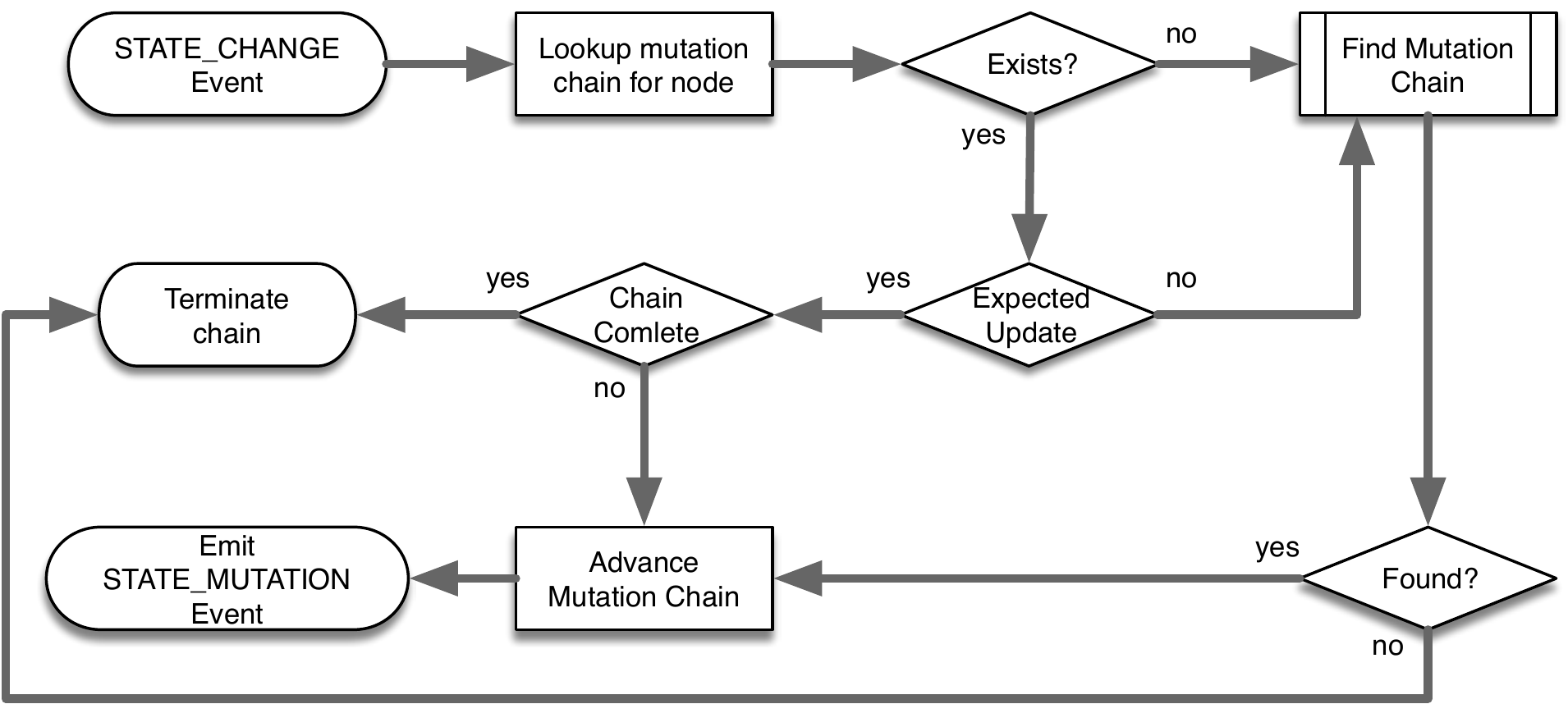}
	\caption{Event handling in the StateMutationEngine.}
	\label{fig:sme-events}
	\end{center}
\end{figure}
\paragraph{StateMutationEngine} The StateMutationEngine (SME) is responsible for finding, executing, and maintaining mutation chains for each node it knows.  For parents, this is generally the node itself and all of its children.  Children only mutate themselves.  The SME is the most complex component in Kraken, and is responsible for meeting most of the requirements.  The SME operates by listening for \prol{STATE_CHANGE} events and responding to them (See Figure \ref{fig:sme-events} for an event flow). The SME can request module execution with \prol{STATE_MUTATION} events.

In addition to handling \prol{STATE_CHANGE} events and emitting \prol{STATE_MUTATION} events, the SME is responsible for providing the unification algorithm to find mutation chains.  We will discuss this separately in Section \ref{subsection:mutations}.  The SME is also responsible for some other minor functions, such as handling mutation failures.  

\paragraph{StateSyncEngine} The StateSyncEngine (SSE) is responsible for synchronizing state across the nodes.  We use a datagram-based eventual-consistency protocol to synchronize state.  We will discuss this in more detail in Section \ref{subsection:state-sync}.

The SSE is also responsible for tracking the availability of nodes.  When a child Kraken instance is started, after basic initialization, its first action is to ``phone home'' to the parent.  This operation establishes state synchronization between child and parent.  The SSE is the only internal component in Kraken that declares a mutation\footnote{For this, and similar reasons, the SSE may eventually be converted into a core module.  This would also allow easy plug-in of different state sync methods.}.   This mutation tracks the \prol{RunState} of a node.  A \prol{RunState} of \prol{SYNC} indicates that state is being synchronize.  Should a node fail to synchronize past a ``dead timer'' it will go into a failed state, which may be recoverable depending on the automation tool.

\paragraph{ServiceManager \& ModuleAPI} The ServiceManager (SM) and the ModuleAPI (API) are two distinct components, but they operate closely together.  The two, together, handle most of the work of providing modularity to Kraken (R\ref{req:mod-mut}-R\ref{req:mod-disc}).  Modules within a Kraken tool run within a different process space.  This is achieved by re-executing the binary, similar to a ``BusyBox'' style of execution.  Modules may have more than one running instance; we refer to these instances as Service Instances.  

The SM is responsible for tracking the desired states of services, running services, tracking their process states, and reporting on the service states.  Services can be requested to be in particular states either by explicit requests in the \prol{Node} state or by the SME to satisfy the need for a mutation.

Once a module is running, it must maintain communication with the core Kraken instance.  Module communication is handled through gRPC communication over a UNIX domain socket.  The API handles these gRPC communications.  It provides event streams for events that the module will need (e.g.~\prol{STATE_MUTATION} events, or events the module subscribes to). The API provides a mechanism for declaring state discoveries and emitting other events.  It also offers essential functions like access to the SDE query engine and message logging through the primary logging mechanism.

\begin{figure}[htb]
	\begin{center}
	\includegraphics[width=\columnwidth,angle=0]{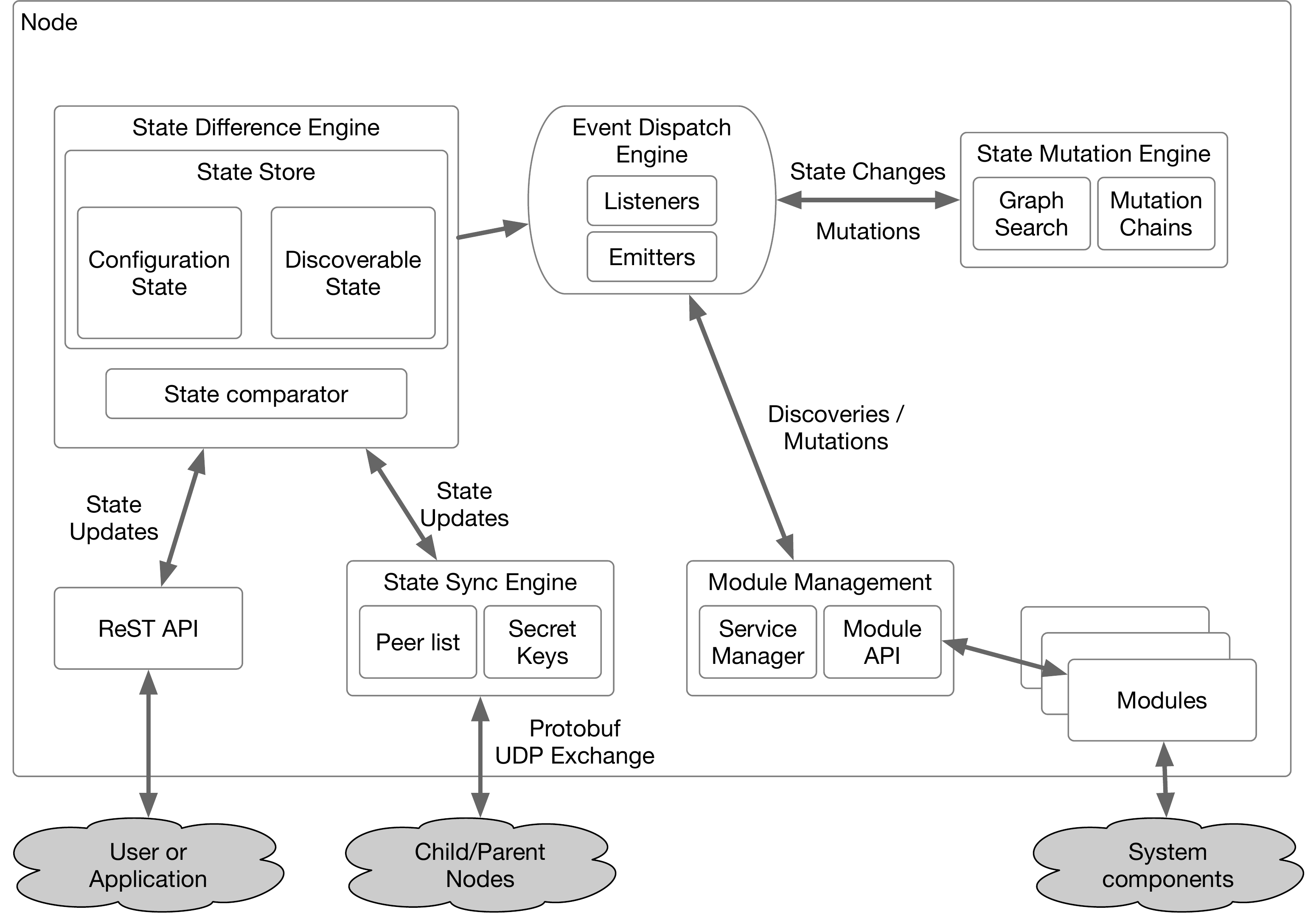}
	\caption{Kraken simplified data flow schematic.}
	\label{fig:kraken-dataflow}
	\end{center}
\end{figure}
\paragraph{Kraken data flow}

Having described the Kraken engines, we present a basic schematic of how the core pieces of Kraken are interconnected (See Figure \ref{fig:kraken-dataflow}).  In this diagram, the arrows represent flow of information (events or queries).

\subsection{State mutations \& unification}
\label{subsection:mutations}

The unification algorithm, as described in Section \ref{subsubsection:state-convergence}, has two significant drawbacks when unconstrained.  First, this algorithm is not guaranteed to finish in finite time without restrictions on what states and mutations are allowable; \emph{a priori} there is nothing that prevents an infinite depth search.  Second, note that even if the algorithm does complete--say, by putting a limit on search depth--each step of the search unifies with each mutation, making this a costly algorithm.  If we limit search depth to $d$ recursions, this is an $O(N^d)$ algorithm.  For practicality and scalability, we must put constraints on the search space.

We will show that the following set of constraints can lead to an efficient algorithm that completes in finite time.  First, we introduce a term that will appear throughout: a \emph{mutation variable} refers to a state variable that the provided set of mutations can mutate.  Mutation variables may make up a small subset of all variables.  In the example represented by Listing \ref{lst:power2}, for instance, \prol{power} is a mutation variable, but \prol{platform} is not; no rules exist to alter \prol{platform}.
\begin{description}
    \item[A1] Every mutation variable must have a finite enumeration.  Specifically, we restrict ourselves to Protobuf \prol{enum} types.
    \item[A2] Every mutation variable must possess a zero value that represents a state of being unknown; the default initial state for every variable is unknown.  It follows that any state variable that appears in a mutation chain must have a mutation available that mutates the variable away from \prol{unknown}.  We call this the \emph{discovery mutation}.
    \item[A3] A state that does not meet the requirements (implicitly or directly) for the discovery mutation of a variable must possess an unknown value for that variable. 
\end{description}
We will demonstrate how these constraints lead to the construction of an optimized graph that can be used to find mutation chains efficiently.  We refer to this graph as an \emph{epistemic state graph}, for reasons that will become clear as we explore them.

\subsubsection{Epistemic state graphs}
\label{subsubsection:epistemic-state-graphs}
\begin{figure}[htb]
	\begin{center}
	\includegraphics[width=\columnwidth,angle=0]{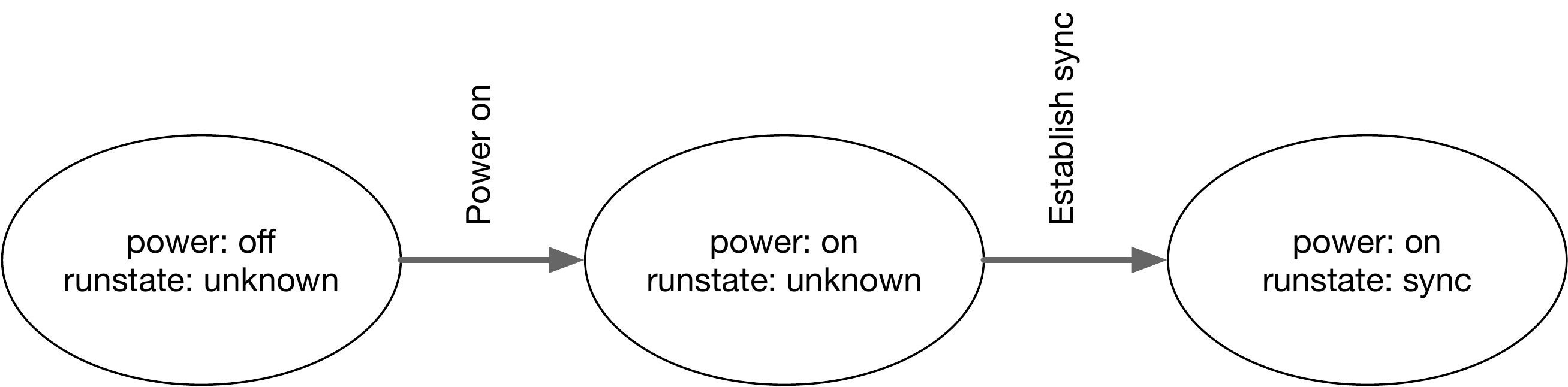}
	\caption{Path representation of a simple mutation chain where nodes are states and edges are associated with mutations.}
	\label{fig:simple-chain}
	\end{center}
\end{figure}
Without applying any of the assumptions above, we make the following observation: a mutation chain naturally forms a directed path.  Since the path is a sequence of mutations of state, we can visualize the nodes of the path as states, and each edge has a mutation associated with it (See Figure \ref{fig:simple-chain}). Note that the mutations need not be unique within the path.  

We can make the following lemma:
\begin{lemma}
    \label{lemma:unique-nodes}
    No state can appear twice in a mutation chain.  Suppose it does.  This implies that two unifications lead to the same state, but by step (2b) of the unification algorithm, this would have terminated the path search.
\end{lemma}
Without this lemma, we would have to assume that different nodes of the same state exist in the same graph.  With this lemma, it is safe to assert an equivalence relation for nodes: \emph{nodes are equal if their states are equal}.  Let's also assign an equivalence relation for edges: \emph{edges are equal if they join the same nodes and they are associated with the same mutation}.  With the reasonable (though not necessary) restriction that we disallow mutations that perform the same mutation and have the same requirements, this simplifies to: \emph{edges are equal if they join the same nodes}.

Consider a graph that is formed from the union of two mutation paths, that is $G(V(1),E(1))\cup G(V(2),E(2)) := G(V(1)\cup V(2), E(1)\cup E(2))$, where we follow the notational convention that $G(V,E)$ is a graph comprised of nodes (vertices) $V$ and edges (arcs) $E$.  We can see that the resulting graph will be simple since mutation chains contain no loops, and edges between the same nodes are equivalent; hence there will be no parallel edges (assuming the stronger equivalence relation for edges).  The graph will be connected if the sets of nodes are not disjoint.

Let $M$ be the set of all mutation paths.  We can define the state graph: $S := \bigcup\limits_{m \in M} m$ .
That is, $S$ is the union of all mutation paths. From the above, we know that this is a simple, directed graph.  It may or may not be connected.  We can also infer that, given $S$, a path search in $S$ between a discovered and configured state is equivalent to the unification algorithm up to path redundancy.  We can search this graph to obtain valid mutation paths, though if there are multiple valid paths, we are not guaranteed the same path.

The state graph $S$ along with an adequate graph search provides us with a new unification algorithm that can perform with graph search characteristics.  For instance, by employing Dijkstra's algorithm\cite{Dijkstra_1959}, we can achieve a performance characteristic of $\Theta(|V|^2)$.  We can also be sure to find \emph{shortest} paths, which the unification algorithm does not guarantee.  

We have yet to apply any of our assumptions to this graph at this point, and it should be noted that $|V|$ could be infinite.  The graph $S$ is currently only generically useful on a computer with infinite memory (and time).  Let's apply (A1) and enforce that mutation variables are finite and enumerable.  With this assumption, we can write another simple lemma:
\begin{lemma}
    \label{lemma:finite-chains}
    If all mutation variables are finite and enumerable, all mutation chains are finite.  Since the cartesian product of all sets of state values, which produces the set of all possible states, is also finite. and given lemma \ref{lemma:unique-nodes}, nodes cannot repeat in a chain.  The longest mutation chain possible is, therefore, equal to the order of the set of all states, which is finite.
\end{lemma}
It follows immediately that the state graph, $S$, under this condition is also finite.  Subsequently, the graph can be stored in finite memory, and the search can be completed in finite time.

To construct this graph algorithmically, we would need to perform a deduction for every possible state combination, collect each path, then perform the graph union to arrive at $S$.  Not only is this approach costly, but it would also create ``unreal'' states.  For instance, using the same states from before, consider the state consisting of: \prol{power = off, runstate = sync}.  This is a valid state combination, but practically this doesn't make any sense.  How can a node that is powered off be synchronizing state?

\begin{figure}[htb]
	\begin{center}
	\includegraphics[width=\columnwidth,angle=0]{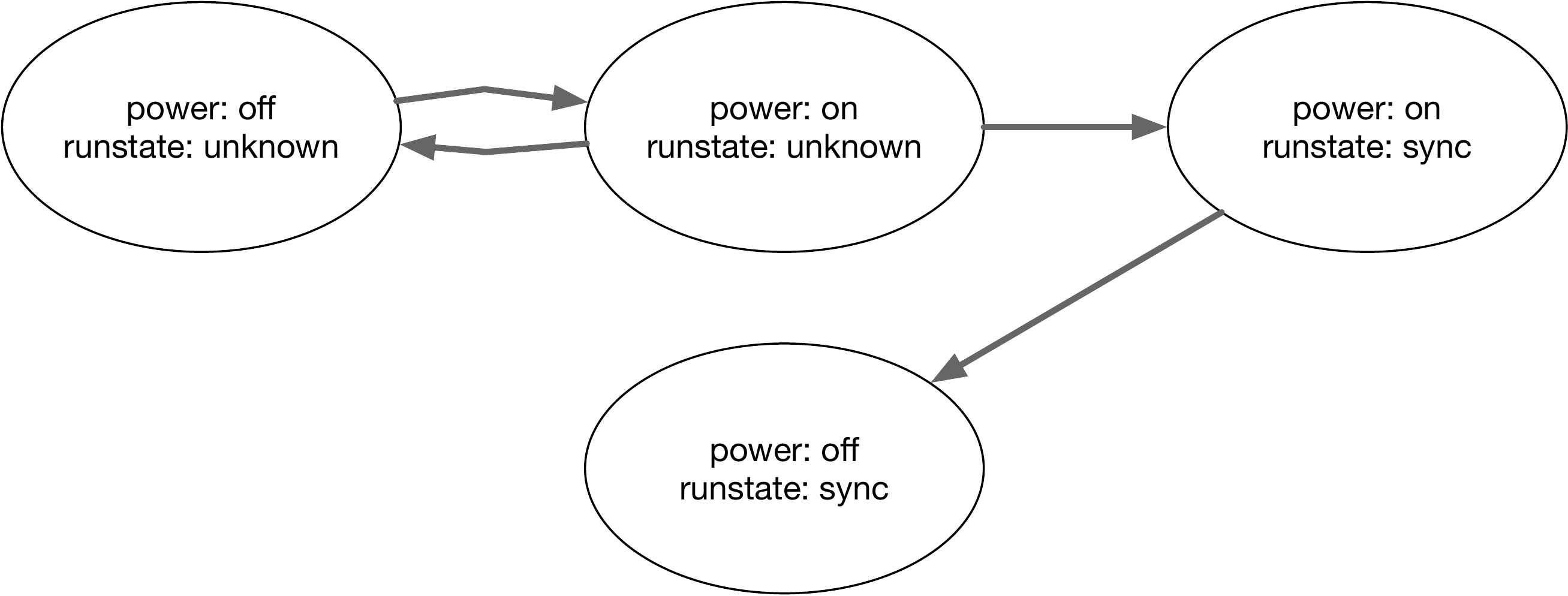}
	\caption{A simple state graph with ``unreal'' nodes.}
	\label{fig:no-epistemology}
	\end{center}
\end{figure}
Even worse, on the union of the paths, these nodes can become connected.  Consider Figure \ref{fig:no-epistemology}.  We can imagine that after reaching a \prol{sync} state and powering off, we would end up in a node with this state configuration.  There are at least two problems here: 1) this node is a sink in the graph, and there is no way to escape it; 2) this node clearly doesn't represent anything real for the presumed system.  

Programmatically, this state is valid.  We need a general way to induce states like this to be logically incorrect.  We could attempt to make rules like: \emph{if the power is off, the runstate must be not-syncing}.  More generally, we can observe that if the power is off, the \prol{runstate} is in a state that we cannot observe (or discover).  Properly speaking, this means the value is unknowable.  Extrapolating, we can see that on startup, all (discoverable) state values are unknown.  This implies the need for the assertion (A1) that every state variable must have an \prol{unknown} value. Moreover, this is its default state until a \emph{discovery mutation} is executed, which discovers the state value.

\begin{figure}[htb]
	\begin{center}
	\includegraphics[width=\columnwidth,angle=0]{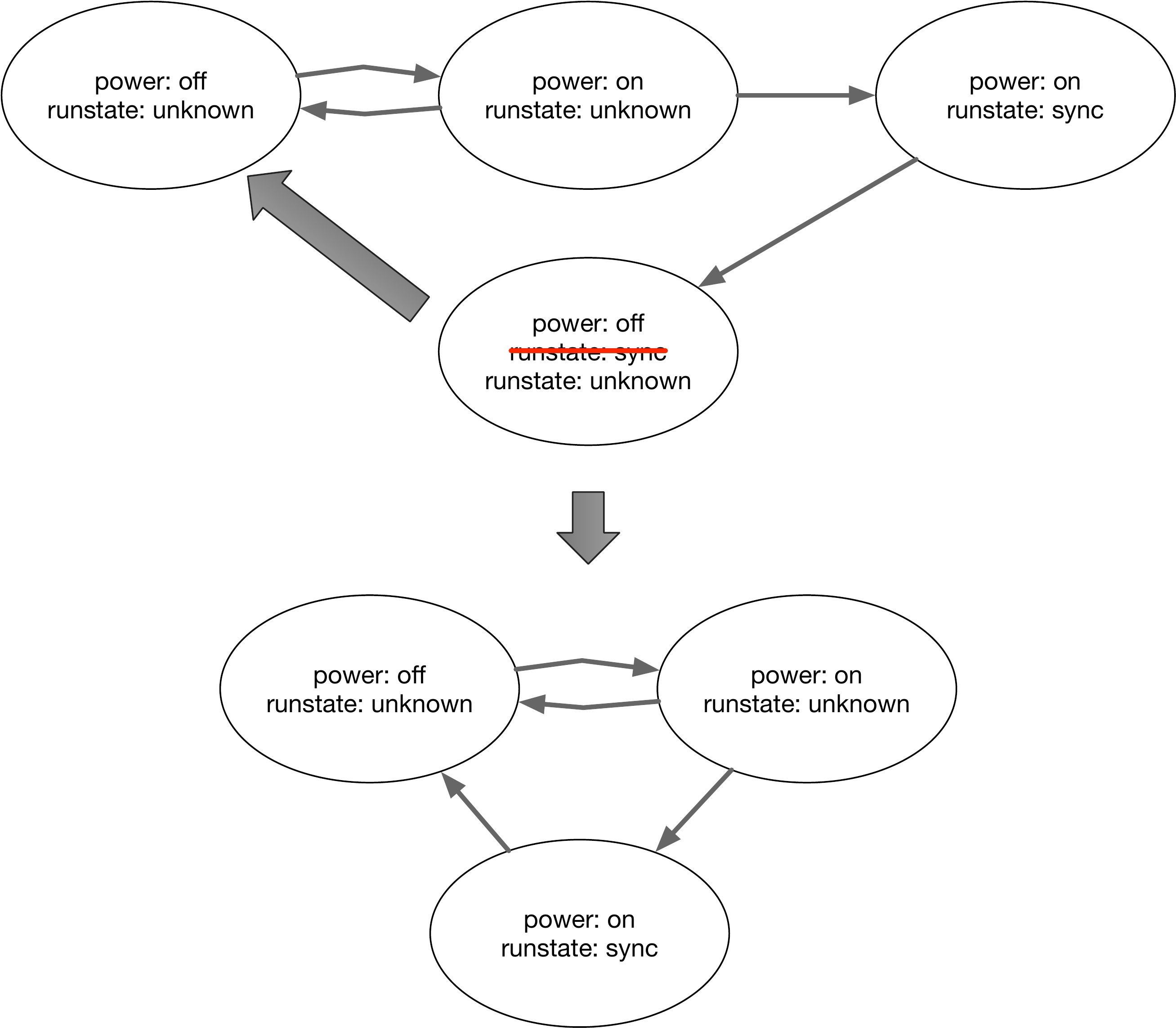}
	\caption{An illustration of epistemic stripping. On stripping the epistemic violation, ``unreal'' node merges with a ``real'' node. }
	\label{fig:graph-epistemology}
	\end{center}
\end{figure}
This assumption allows us to say \emph{why} the ``unreal'' node above is wrong.  Notice that the mutation that changes \prol{runstate = unknown -> runstate = sync} is a discovery mutation.  We can infer that a requirement of that mutation (based on its placement in the graph) was \prol{power = on}.  When we move to a state \prol{power = off}, the requirements for the discovery mutation are no longer met.  We can resolve this by implementing assumption (A3): when a state no longer meets the requirements for the discovery of a variable, that variable becomes \prol{unknonw}.  

Looking at Figure \ref{fig:graph-epistemology}, we see that the runstate in the ``unreal'' node becomes unkown.  This results in a merge with th existing start node, making a more ``real'' graph.
We refer to this process as \emph{epistemic stripping} because it tracks what values are knowable and strips unknowable values.  We refer to graphs that adhere to this principle as \emph{epistemic state graphs}.

\subsubsection{Graph building \& searching}
\label{subsubsection:graph-building}
The Kraken framework uses epistemic state graphs based on the mutations provided by modules to efficiently provide mutation chain lookups for Kraken-based tools.  This is made possible under the assumptions (A1-A3) above.  Kraken does not, however, construct graphs in the manner described above.  This approach is far more costly than necessary.

Notice that assumption (A2) implies that every graph contains a node in which all mutation variables are \prol{unknown}.  We can construct an epistemic state graph by constructing a graph that searches for all paths connected to the root.  We do this in three phases: 

\paragraph{Phase 1} We define a mutation as being \emph{out-wise compatible} with a state if the state does not conflict with the requirements of the mutation.  We define a mutation as being \emph{in-wise compatible} with a mutation if the node that would be implied by rewinding the states mutated by the mutation is compatible with the mutation. Starting from the root (all-unknown) node, we can find all in-wise and out-wise compatible mutations.  We then create an edge for each mutation, pointing in or out depending on in-wise or out-wise compatibility.  Finally, we construct a new node with the state that is implied by playing forward or rewinding the mutation.  If a node has already been seen, we map the edge to that node.  We proceed in this way recursively until all nodes have been seen.  At the end of Phase 1, we have an expanded graph, but it has not accounted for epistemology.

\paragraph{Phase 2} In this phase, we derive an epistemology.  The epistemology is a record of what states must exist for a variable to be knowable.  We derive this by searching the Phase 1 graph for any instances of discovery mutations.  We then perform an exclusive merge on all of the nodes on the tail of the edge.  That is, we keep any state value that is common to all such nodes.  The remaining values after the merge we count as being required to know the variable in question.  This constitutes our epistemology.

\paragraph{Phase 3} In the final phase, we apply the epistemology derived in Phase 2 to the graph derived in Phase 1.  We iterate through the nodes of the graph, epistemically stripping them.  We then collapse any equivalent nodes and strip any orphaned or equivalent edges.  

This process produces a functional epistemic state graph.  Notice that Phases 2 and 3 are identical to how we described the purely mathematical approach to graph construction.  All that needs to be demonstrated to show the equivalanece of this graph with the mathematically constructed graph is to demonstrate equivalance of the Phase 1 graph to the non-epistemically stripped state graph.  Since Phase 1 explores all possible mutations connected to the root node, we conclude that Phase 1 is equivalent \emph{up to connectedness to the root node}.  Noting that this is an implied requirement for states to be known, and given A1 and A2, we conclude that these two graphs are \emph{functionally} equivalent, the algorithmically assembled graph may differ in sets of disconnected, unreachable subgraphs.

With the graph available, we can rapidly find mutation paths by finding endpoint nodes and performing a graph search.
All of the graph building and searching functions are performed by the StateMutationEngine, and are linked to the event-driven design in the way outlined in Section \ref{subsectinon:architecture-overview}.

\subsection{State sync \& eventual consistency}
\label{subsection:state-sync}
State synchronization is another area where the Kraken framework makes optimizing assumptions.  State synchronization can significantly impact scalability, depending on the implementation.  Two factors tend to drive scalability concerns for state synchronization: 1) the consistency guarantees of the synchronization; and 2) the volume of data to be synchronized.

Within Kraken, we optimize state synchronization with the following two additional assumptions:
\begin{description}
    \item[A4] We will never guarantee consistency of state across all nodes, but, in the absence of further changes, it will converge within a well-defined time.
    \item[A5] Aggregate state information for a single node should be small relative to a network transport frame.
\end{description}
These two assumptions address the two state synchronization scalability concerns, respectively.

Consistency of state in distributed systems is a widely studied field. Consistency is classed into different levels based on guarantees surrounding when a parallel process will see the updates from another process.  Strong consistency, for instance, is the case in which all access to data from all processes is enforced to be sequential.  This means that there will never be a time when a value is changed, but a read from another process sees an old value.

Guarantees like those provided by strong consistency can come at a high performance cost\cite{terry}.  For instance, well-known algorithms such as Paxos\cite{paxos} and Raft\cite{raft} provide strong consistency across distributed systems using 2-phase commits.  However, the consensus gathering in these algorithms grows in complexity rapidly with the number of systems.  For this reason, platforms implementing these, like Zookeeper\cite{zookeeper} and etcd\cite{etcd} are intended to run as small clusters that provide state query services to a large pool of systems. This model does not meet our needs for state synchronization.  Each automation actor needs its piece of the state.  We contend that we can live with weaker consistency guarantees for a large class of systems automation problems (see Section \ref{section:problem-constraints}).

The extreme end of the spectrum from strong consistency is no consistency.  That is, we make no guarantees whatsoever about the consistency of state data.  Eventual consistency\cite{vogels} is a slight hedge from no-consistency.  It also makes no guarantees about the consistency of data at any given moment, but it does guarantee that, in the absence of further changes, the state will converge to consistency within a defined time window.  Eventual consistency chooses availability and partition tolerance over consistency\cite{captheorem,terry,vogels}.  More importantly for us, it can be highly performant and scalable.

When considering state propagation, it is always important to consider the source of truth for the state.  This is especially true when dealing with an eventually consistent propagation.  In general, the processes closest to the source of truth can be better trusted.  In the Kraken model, the node's discoverable state is closest to the node itself, while the configuration state is closest to the node's parent.  Once a node begins synchronizing state, we ``trust'' the discoverable state from the node over the parent's version, and conversely, we trust the configuration state from the parent over the nodes' local copies.

This flow of state represents how we establish state synchronization within Kraken.  Once a ``phone home'' process has taken place (a gRPC call), the node and parent start a synchronization scheme that is similar to how routing protocols such as OSPF and IS-IS synchronize routing information.  On a regular ``hello'' timer, each node sends a state synchronization packet to its neighbors.  From parent to child, this contains the configuration state relevant to the child.  Conversely, the child sends its parent its discoverable state.  In this way, state information flows from its respective source of truth.  If a synchronization packet is not received within a ``dead'' timer window, the neighbor is considered to have died. 

By implementing this protocol with one-way datagram packets, we can achieve robust scalability for state synchronization.  Implementing using a simple datagram protocol also implies the assumption (A5): we are driven to keep the size of state synchronized very small.  By doing so, we gain additional advantages, though.  For systems that synchronize large amounts of data, sophisticated, load-balanced mechanisms are often necessary for data transport (e.g.~\cite{arnold2006tree}).  By enforcing total state information to be small, we can avoid these complexities.

%% file: 5.0-problem-constraints.tex

Overall, the Kraken framework implements the requirements outlined in Section \ref{section:anatomy}, and can be used to design a wide variety of distributed automation workflows that meet the stated design objectives.
As part of the implementation, we noted five optimizing assumptions.  Each of these imposes some limitations on the problems that we can solve using the Kraken framework. We address these now.

\subsection{Kraken framework limitations}

A1 stated that we can only use finite enumeration types as mutation variables.  Initially, this appears to be limiting.  This means, for instance, that you cannot declare a mutation that mutates a string.  However, there are generally workarounds for these issues in practice, and often the end result is more programmatically sound. Normally, if a workflow needs to modify some non-enumerable value, it can do so by being paired with an enumerable value.  This value can track when the non-enumerable value needs an update, note the need for an update as within the enumerable value, and Kraken can inform the module to perform the required update when all requirements for doing so are met.
This method can extend to much more complex cases, such as tracking complex sets of non-enumerable variables.  

A2-A3 state that each enumerable needs an ``unknown'' value and that this is the default initialized value.  Additionally, the need for a discovery mutation is implied.  These constraints do not impose any legitament limitations on the automation flow.  In most cases, the idea that mutation variables start as unknown is natural.

A4 establishes the lack of a consistency guarantee on the state.  This means that a Kraken tool might act on the wrong mutation chain for a short period of time.  This can impact which actions should be allowable in a mutation chain.  As long as mutations are correctable, this is not a significant concern.  For instance, if an out-of-sync piece of configuration data causes a node to begin loading an incorrect image, this can be (in most cases) reversed and corrected.  We are driven to adopt a model that is familiar in, for instance, many areas of container orchestration: failures are fine as long as they get corrected (quickly).

There are certain operations for which Kraken would not be a good choice.  In general, issues, where even temporary failures in data integrity are non-reversible or catastrophic, are bad use-cases for Kraken.  However, in some cases, Kraken can be combined with other methods or tools to provide stronger consistency in isolated cases.  

A5 requires aggregate state data for nodes to be small.  We have not yet seen a case where this is constraining to automation logic.  It can be tempting to include, e.g.~large amounts of configuration data in the state.  This is generally not well supported due to this constraint, and large data blobs should be acquired from outside of state synchronization.

\subsection{Use cases}
Having covered some of the areas where Kraken has limitations, we now briefly cover promising areas for Kraken tooling.  We begin with some projects that have already been explored:
\begin{description}
    \item[HPC Provisioning] Kraken was originally conceived to implement more powerful HPC provisioning and node lifecycle management tools.  As such, the flagship Kraken-based tool is the ``Layercake'' tool.  We will take a closer look at this case below.
    \item[Power \& Thermal scaling] The automation capabilities of Kraken could make a powerful set of tools for scaling power and thermal consumption dynamically.  A proof-of-concept of this has been implemented that can regulate power consumption and heat output of a cluster by dynamically adjusting CPU frequency scaling policies across the cluster.
    \item[Controlling other API-driven tools] In some cases, an existing platform exposes enough through APIs that Kraken can act as an orchestrator while not directly performing tasks.  An existing proof-of-concept of this approach uses a Kraken tool to manage node lifecycle and provide robust, state-aware rolling-update capabilities to Cray Shasta systems. 
    \item[Remote application management] Managing the running state of applications or containers across a cluster has been proven out through different projects.  This allows a Kraken tool to act as a kind of application orchestrator across a large set of systems. 
\end{description}

We believe there are many possibilities for the Kraken framework that have not yet been explored.  Kraken should perform well for any project that can tolerate eventual consistency and can benefit from declarative management and continuous state enforcement.

\subsection{Case study: Kraken/Layercake}
The flagship Kraken-based tool is the Layercake tooling\cite{layercake} for cluster provisioning and node lifecycle management.
``Layercake'' refers to the model used to manage system images, which follows the pattern outlined in \cite{layercake-paper}. As opposed to traditional HPC management systems, Layercake provisions nodes in layers.  It first brings a system from cold-boot to running a persistent minimal OS (layer 0).  This OS provides a very lightweight environment that includes, among other things, a running layercake instance.  From this point, the Kraken-based automation will load container sets as system images.  Because the minimal OS is persistent, layercake can ``roll'' images rapidly, often in under a second.

The benefits of having Layercake based on Kraken comes out in three primary ways. First, owing to Kraken's modularity, Layercake can natively support \emph{many} ways to bring systems to desired states.  This includes everything ranging from different power control mechanisms to completely different boot processes.  If a sequence of mutations can be provided to get a system to the needed states, modules can be built for layercake to support the system.  This provides Layercake a lot of flexibility that would otherwise be very difficult and likely very manual.

Second, because everything Layercake does is state-aware, it can seamlessly provide state-aware workflows.  For instance, by excluding ``busy'' states for operations that would interrupt a running system image, we can declare a new image for all of the nodes in a cluster, and they will ``roll'' to the new image as soon as each node is free, allowing for seamless image updates that do not interrupt running work.

Finally, because Layercake uses Kraken's continuous automation, a Layercake cluster is capable of some self-healing.  For instance, errors in the PXE boot process, or a crashed image (container) can be automatically recovered.

As an added benefit, because much of the internal logic of Layercake is handled by Kraken, the coding effort for the Layercake modules is minimal.

%% file: 6.0-conclusion.tex

Starting from a base set of requirements, we have prescribed a theoretical model for automation tools that are distributed, scalable, modular, and continuous.  While we make no claim that this model is unique in its requirements, we believe the requirements can be a rubric for current and future projects that wish to meet these objectives.

We have also presented the Kraken framework, which facilitates the creation of tools that meet these design objectives under certain constraints.  We believe the ability to create these tools efficiently can fill many gaps in currently available tooling.   We also believe that some of the methods used in designing the Kraken framework, such as epistemic state graphs and eventual consistency for automation, will prove useful in other contexts.

The Kraken framework, and some of the derivative tooling, while still in relatively early development stages, have already proven capable of delivering innovative management capabilities for distributed systems.  We believe that we have only begun to scratch the surface of what is possible using these tools.

%% file: reqs.tex
\section{Table of requirements}
\setcounter{reqs}{1}
\begin{description}
    \req{state-mgmt}{State management}{We must possess a representation of the current state of the system along with the tooling to track, evaluate and report on these states.}
    \req{state-spec}{State specifications}{We must possess a mechanism for defining and asserting goals, or desired states,  for the system.}
    \req{state-conv}{State convergence}{We must possess a mechanism to find and actuate a sequence of state mutations (a \emph{mutation chain}) that will achieve the desired state.}
    \req{discoverable}{Discoverable}{States which have a source of truth outside of the program must have a means of inspecting and updating the local representation of the states.}
    \req{event-driven}{Event-driven}{To achieve continuous operation and reliable execution, changes in discoverable state or state goals should be handled as asynchronous events.}
    \req{cfg-states}{Configuration states}{We must provide a mechanism or API for users or programs to set and update configuration states. Note that this is related to but distinct from (R\ref{req:state-spec}).}
    \req{mutations}{Mutation management}{We must provide a mechanism for finding and executing chains of state mutations.  Note that this is related to but distinct from (R\ref{req:state-conv})}.
    \req{sync}{State synchronization}{To distribute automation, we must provide a mechanism to synchronize state across system nodes.}
    \req{local-execution}{Local execution}{To distribute automation, we must provide a local automation execution environment on nodes.}
    \req{mod-mut}{Module mutations}{To achieve modularity, state mutations should be provided by modules.}
    \req{mod-disc}{Module discoveries}{To achieve modularity, discoverable states should be assigned by modules.}
\end{description}